\documentclass[aps,twocolumn,superscriptaddress]{revtex4}
\usepackage{amssymb}
\usepackage{graphicx}

\input{tcilatex}

\begin{document}

\title{Effects of Finite Deformed Length in Carbon Nanotubes}
\author{Jun-Qiang Lu}
\affiliation{Center for Advanced Study and Department of Physics, Tsinghua University,
Beijing 100084, China}
\author{Jian Wu}
\affiliation{Center for Advanced Study and Department of Physics, Tsinghua University,
Beijing 100084, China}
\author{Wenhui Duan}
\affiliation{Center for Advanced Study and Department of Physics, Tsinghua University,
Beijing 100084, China}
\author{Bing-Lin Gu}
\affiliation{Center for Advanced Study and Department of Physics, Tsinghua University,
Beijing 100084, China}
\date{\today }

\begin{abstract}
The effect of finite deformed length is demonstrated by squashing an
armchair (10,10) single-walled carbon nanotube with two finite tips. Only
when the deformed length is long enough, an effectual
metal-semiconductor-metal heterojunction can be formed in the metallic tube.
The effect of finite deformed length is explained by the quantum tunnelling
effect. Furthermore, some conceptual designs of nanoscale devices are
proposed from the metal-semiconductor-metal heterojunction.
\end{abstract}

\maketitle






The discovery of carbon nanotubes\cite{SI} offers a natural quasi one
dimensional material to explore and verify quantum physics theory, and the
advantage in design of nanodevice. Recently, an interesting topic is how a
mechanical deformation of a single-walled carbon nanotube (SWNT) affects its
electronic property.\cite{TWT,EDM,JC,AM,MBN} Some previous studies\cite%
{AM,MBN} show that in many cases, a local deformation can not drive a
metal-to-semiconductor transition (MST) in armchair SWNTs. At the same time,
our recent work \cite{JQL} indicates that a MST can be achieved by a radial
deformation along the holistic armchair SWNTs. The different conclusions
drawn from the two models show that the finiteness of the deformed length
plays an important role in the transport properties of SWNTs. So, for
designing electronic devices based on nanotubes, the deformed length along
the tube will be a key parameter to ensure their functions. Here, in this
letter, we will demonstrate the effects of the deformed length by squahing
an armchair (10,10) SWNT.

In order to simplify the theoretical analysis, we consider an armchair
(10,10) SWNT symmetrically squashed by two identical tips (with a width of $%
d_{x}=4.0$ \r{A}~ and a length of $d_{z}$) along the $\pm y$ direction till $%
d_{y}=2.2$ \r{A},\cite{JQL} as shown in Figs. 1a and 1b. The atomic
structures, or the shape of tubes, for different tip lengths $d_{z}$, are
optimized by the tight-binding (TB) molecular dynamics (MD) code.\cite%
{CHX,JQL} Fig. 1a shows an example optimized structure for tip length $%
d_{z}=6.0$ \r{A}. Noticeably, the cross section of the tube changes from a
dumbbell (Fig. 1b, corresponding to fully deformed part of the tube) to an
ellipse (Fig. 1c, less deformed part), and then to a circle (Fig. 1d,
undeformed part). The smallest interatomic distance between the upper face
and the lower face is $d_{AA^{\prime }}=2.1$ \r{A}, and the mirror symmetry
is broken due to small $d_{y}$. In this case, the squashed part should work
as semiconductors.\cite{JQL}

\begin{figure}[b]
\begin{center}
\includegraphics[width=2.8in]{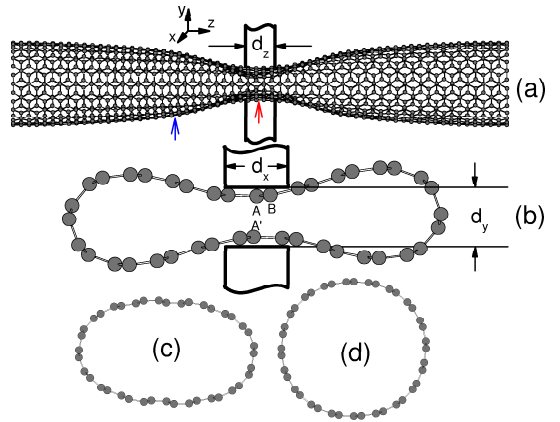}
\end{center}
\caption{(color) (a) Finite length deformation in an armchair (10,10) SWNT
squashed by two identical tips; (b)(c) Cross sections of the different parts
of the deformed SWNT, as indicated by the red and blue arrows, respectively;
(d) Cross section of the non-deformed part of the SWNT.}
\end{figure}

After obtaining the optimized structure with different tip lengths, we
employ the TB Green's function method\cite{JCC,JW,JW2} to study the
electronic transport properties of the deformed tubes. Fig. 2a presents the
conductance curves near the Fermi energy $E_{F}$ of tubes deformed by tips
with different lengths $d_{z}$. We can see that no energy gap is opened when 
$d_{z}$ is $3.0$ nm (the pink short dash line). In other words, the deformed
tube keeps its metallic behavior. Moreover, the conductance of the deformed
tube almost keeps constant near $E_{F}$. Only when the tip length increases,
the conductance near $E_{F}$ will begin to decrease, as shown in Fig. 2a. It
is easy to find that the conductance near $E_{F}$ approaches zero when $%
d_{z}=25.2$ nm (the red dash line). As the limit, when the length of tips
approaches infinite, the whole tube is deformed, and then the conductance
near $E_{F}$ will be zero. 

\begin{figure}[t]
\begin{center}
\includegraphics[width=2.8in]{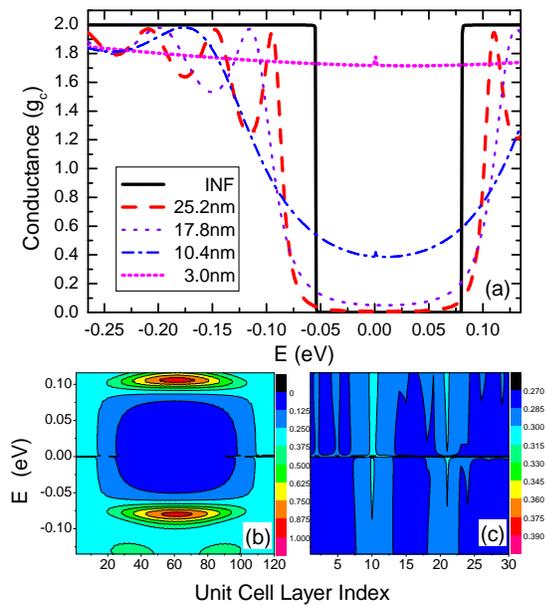} 
\end{center}
\caption{(color) (a) Conductances of SWNTs deformed by tips with different
length $d_{z}$. $g_{c}(=2e^{2}/h)$ is the unit quanta of conductance. $E$ is
the energy of injected electron, and the Fermi energy of ideal armchair
(10,10) SWNT is taken as zero. (b)(c) The LDOS (unit: eV$^{-1}$) per unit
cell layer near $E_{F}$ of SWNTs deformed by tips with lengths (b) $%
d_{z}=25.2$ nm and (c) $d_{z}=3.0$ nm.}
\end{figure}

The above results demonstrate that the deformed length in SWNTs plays an
important role in their electronic properties. In order to reveal the
effects of finite deformed length in SWNTs, we present a visual picture by
the local density of states (LDOS)\cite{JW} per unit cell layer along the
tube axes. Each unit cell layer consists of 40 atoms which distribute over
two atomic layers. When $d_{z}=25.2$ nm, the LDOS of 120 unit cell layers
(4800 atoms) is plotted in Fig. 2b, as the deformed part of the tube
contains about 100 unit cell layers. It is obvious that the LDOS of the
deformed part is almost zero near $E_{F}$. Energy gap exists near $E_{F}$
within the deformed part. While, the LDOS crosses $E_{F}$ continuously in
the nearby part. Consequently, the squashed tube forms a
metal-semiconductor-metal (MSM) heterojunction. Meanwhile, when $d_{z}=3.0$
nm, the LDOS of 30 unit cell layers (1200 atoms) is plotted in Fig. 2c,
since the deformed part only includes about 12 unit cell layers. The LDOS
crosses $E_{F}$ continuously in both the deformed part and its nearby.
Hence, the tube keeps its metallic behavior.

Physically, the effect of finite deformed length can be understood by the
quantum tunnelling effect. It is well known that the tunnelling probability
attenuate exponentially with increasing width of energy barrier. Obviously,
electrons can easily tunnel through the energy barrier if it is narrow
enough. This also occurs for nanoelectronic devices. If the size of the
device is too small, the electron will tunnel through the device, and
consequently, the device can not function well as expected. The size effect
is an essential problem for nanoelectronic devices design, as the tunnelling
effect is universal.

To give a vivid sight of the tunnelling effect in the deformed tube, we
present the LDOS of each atom within one unit cell layer of the dumbbell
part in Figs. 3a and 3b. It is well known that the graphene sheet and hence
the nanotube have two equivalent sublattices, which may be labelled as A and
B sublattices. It has been pointed out that, when a MST is achieved in a
squashed armchair SWNT, the LDOS distributing on the two sublattices are
distinguishable, resulting in a discontinuity energy spectrum near $E_{F}$.%
\cite{JQL} This phenomena can also be found in the dumbbell part of tube
deformed by tips with length $d_{z}=25.2$ nm, as shown in Fig. 3a. For the
case of $d_{z}=3.0$ nm, the LDOS distribution of the dumbbell part is
presented in Fig. 3b. We can find only a slight distinction between the LDOS
of the two sublattices. While the LDOS crosses $E_{F}$ continuously. The
LDOS near $E_{F}$ in the dumbbell part should result from the tunnelling
effect.

\begin{figure}[b]
\begin{center}
\includegraphics[width=3.0in]{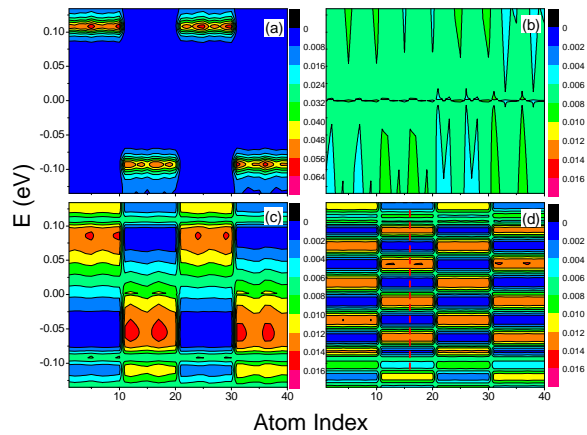}
\end{center}
\caption{(color) The LDOS (unit: eV$^{-1}$) near $E_{F}$ within a unit cell
layer of SWNTs deformed by tips with different lengths (a)(c)(d) $d_{z}=25.2$
nm and (b) $d_{z}=3.0$ nm. (a)(b) dumbbell part; (c) ellipse part and (d)
circle part. The atoms in B (A) sublattice are labelled 1 (11) through 10
(20) for the first atomic layer and 21 (31) through 30 (40) for the second
atomic layer. The dash line is added only for guide.}
\end{figure}

More interesting thing is the change of electronic structure of the tube due
to the finite deformation. The LDOS within a unit cell layer of the ellipse
part (as shown in Fig. 1c) of the tube deformed by tips with $d_{z}=25.2$ nm
is plotted in Fig. 3c. Though there is no energy gap near $E_{F}$ with the
absence of the bond formation between the flattened faces, it is still
interesting to find that the LDOS over the two sublattices are also
distinguishable, which results from the distinction of the two sublattices
in dumbbell part. To reveal this effect, in Fig. 3d, we plot the LDOS within
a circle unit cell layer, which is far from the deformed center with
distance 40.6 nm. To our surprise, again, the LDOS of the two sublattices
are distinguishable. And more strange thing is the LDOS on one certain atom
oscillates with energy, as indicated by the red dash line in Fig. 3d.
Further studies show that the LDOS of the two sublattices within a unit cell
layer are always distinguishable when the distance between the layer and the
deformed center increases, and the oscillation of LDOS with energy becomes
more exquisite. For an ideal SWNT, the LDOS near $E_{F}$ distribute over
each carbon atom homogeneously. With the finite deformation, the $\pi $ and $%
\pi ^{\ast }$ wave functions in the tube are mixed due to the breaking of
the mirror symmetry.\cite{JQL} At the same time, as the translation symmetry
along the tube is also broken, the standing wave functions are the
combination of the incoming and reflecting wave functions,\cite{JW2} and the
LDOS\ of them can be expressed as $2[1\pm \cos (2\Delta kx)][1+R\pm 2\sqrt{R}%
\cos (2k_{F}^{^{\prime }}x+\delta )]$.\cite{HFS} Obviously, the distinction
between the LDOS of the two sublattices results from the difference between $%
1\pm \cos (2\Delta kx)$. For a certain $x$, the oscillation of LDOS with
energy results from $\cos (2\Delta kx)$. As the energy dispersion relations
near $E_{F}$ is approximately linear with $\Delta k=2E/(\sqrt{3}a\gamma _{0})
$, the period of the oscillation is $T_{E}=\sqrt{3}a\gamma _{0}\pi /(2x)$.

From the above discussion, we know that an armchair (10,10) SWNT can form a
MSM heterojunction when the deformation length is large enough ($d_{z}\sim 20
$nm). For the conducting electrons, the middle semiconductor layers act as a
energy barrier. Thus, based on the MSM heterojunction, we can easily
structure some SWNT-based nanoscale devices.

\textit{Double barrier }-- To structure nanoscale double barrier, we only
need to deform an armchair SWNT at two different positions, as shown
schematically in Fig. 4a. As well known, the typical behavior of double
barrier structure is the resonant tunnelling. We find that the conductance
exhibits such a behavior (see Fig. 4a). The small conductance peaks lying in
the energy gap are due to the resonant tunnelling effect. And, it is easy to
control the electronic transport property of the nanoscale double barrier
through changing the widths of the two energy barriers and the middle energy
trap (i.e., changing the deformed length and the distance between the two
deformed positions). For the example case shown in Fig. 4a , the tip length $%
d_{z}=25.2$ nm, and the distance between the two deformed positions $l=57.1$
nm.

\textit{Superlattice} -- If an armchair SWNT is deformed periodically along
its axis, as shown schematically in Fig. 4b, it will behave as a
superlattice. The conductance of the nanoscale superlattice is also
presented in Fig. 4b. The period of the superlattice is $l=57.1$ nm, and the
tip length is $d_{z}=25.2$ nm. Each conductance peak is split into many
small peaks, as shown in the inset. This is the feature of superlattice
structure, named as resonant splitting effect. It is also found that the
conductance vanishes near $E_{F}$ between the conductance bands. We can
control the electronic transport through changing the range of the
conductance gap, which can be achieved by adjusting the structure parameters.

\begin{figure}[t]
\begin{center}
\includegraphics[width=2.8in]{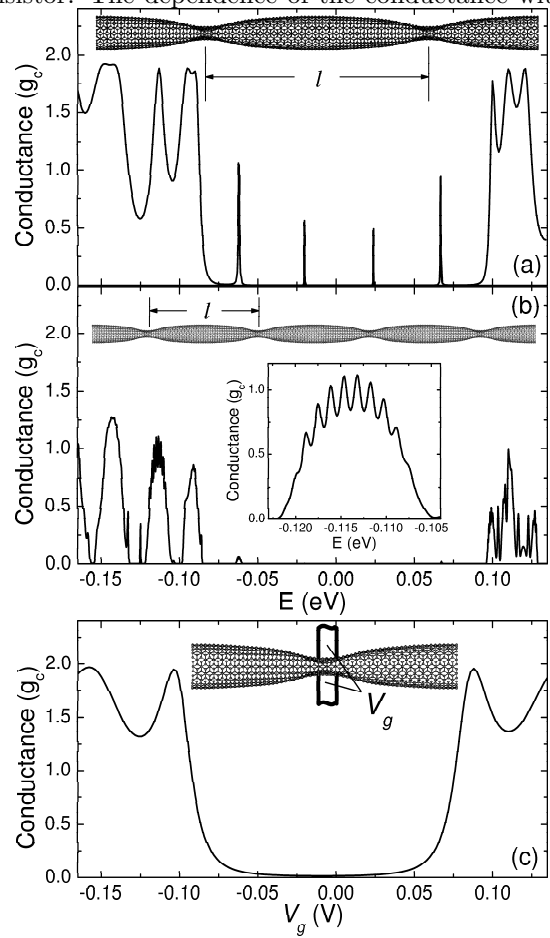}
\end{center}
\caption{Conceptual designs of nanoscale (a) double barrier; (b)
superlattice; (c) transistor and their conductance curves. Note that the
structures are shown schematically only. The inset presents the detail of
resonant splitting effect.}
\end{figure}

\textit{Transistor} -- For a logical circuit, transistor is one of the most
important components. Designing a nanoscale transistor will be one of key
steps to realize the nanologic. Recently, Bachtold \textit{et al}. achieved
a nanoscale transistor experimentally.\cite{AB} They adopted a
semiconducting SWNT. The SWNT needs to be connected with two gold
electrodes. Here, based on the MSM heterojunction, we can design pure carbon
nanoscale transistor.

Fig. 4c is the schematics of a conceptual design of nanoscale transistor ($%
d_{z}=25.2$ nm). If a gate voltage $V_{g}$ is applied to the tips, the MSM
heterojunction will act as a nanoscale transistor. When $V_{g}=0$, as the
deformed part behaves as semiconductor, the conductance is zero and the
transistor is OFF. With the change of $V_{g}$, the energy gap of the
deformed part shifts. When $E_{F}$ does not lie in the energy gap any
longer, the transistor begins to be ON. That is the basic principle of our
nanoscale transistor. The dependence of the conductance with $V_{g}$ is also
presented in Fig. 4c. We can find the logical voltage of the transistor is
about $-0.10$ V.

In summary, we demonstrate the effect of finite deformed length through
squashing an armchair (10,10) SWNT. Only a deformation with enough long
length can lead to an effectual MSM heterojunction in the metallic tube. The
effect of finite deformed length is a general problem and is explained by
the quantum tunneling effect. The change of the electronic structure, or the
LDOS distributions, due to the finite deformation, shows the correlation
between the nanodevices and leads. It should be considered in design of
nanologic. Furthermore, based on the MSM heterojunction, some conceptual
designs of nanoscale devices are proposed.

The work is supported by the National High Technology Research and
Development Program of China (Grant No. 2002AA311153), the Ministry of
Education of China, and the National Natural Science Foundation of China.



\end{document}